\documentclass[twocolumn]{aastex631}

\usepackage{bm}
\usepackage{amsmath}

\usepackage{xcolor}

\newcommand{\Msun}{\text{M}_{\odot}}
\newcommand{\kmsec}{\text{km}\,\text{s}^{-1}}

\newcommand{\kpc}{\text{kpc}}
\newcommand{\pc}{\text{pc}} 

\newcommand{\Myr}{\text{Myr}}
\newcommand{\Msunppcc}{\Msun\pc^{-3}}

\newcommand{\data}{\bm{d}}
\newcommand{\cdata}{\bm{t}}
\newcommand{\simp}{\bm{\theta}}

\newcommand{\update}[1]{#1}

\begin{document}

\title{Using simulation based inference on tidally perturbed dwarf galaxies: the dynamics of NGC205}

\author[0000-0001-5686-3743]{Axel Widmark}
\affiliation{Columbia University, 116th and Broadway, \\ New York, NY 10027 USA}
\affiliation{Stockholm University and The Oskar Klein Centre for Cosmoparticle Physics, \\
Alba Nova, 10691 Stockholm, Sweden}

\author[0000-0001-6244-6727]{Kathryn V. Johnston}
\affiliation{Columbia University, 116th and Broadway, \\ New York, NY 10027 USA}

\begin{abstract}

We develop a novel approach to performing precision inference on tidally perturbed dwarf galaxies. We use a Bayesian inference framework of implicit likelihood inference, previously applied mainly in the field of cosmology, based on forward simulation, data compression, and likelihood emulation with neural density estimators. We consider the case of NGC205, a satellite of M31. NGC205 exhibits an S-shape in the mean line-of-sight velocity along its semi-major spatial axis, suggestive of tidal perturbation. We demonstrate that this velocity profile can be qualitatively reproduced even if NGC205 was in a spherically symmetric and isotropic state before its most recent pericenter passage. We apply our inference method to mock data and show that the precise shape of a perturbed satellite's sky-projected internal velocity field, mapped across its entire face, can be highly informative of both its orbit and total mass density profile, even in the absence of proper motion information. For NGC205 specifically, our method is hampered because the available data only covers a line along its semi-major axis. This shortcoming could be addressed with another round of observations.


\end{abstract}



\section{Introduction}
\label{sec:intro}

The field of galactic dynamics in general, and dynamical mass measurements in particular, is often studied under the assumption of a steady state \citep{Binney2008,Read2014,deSalas2021}. A steady state approximation is powerful as it provides a direct connection between the phase-space distribution of some tracer population and the gravitational potential it inhabits. Parametric models of the stellar distribution and potential can be formulated analytically 
and data likelihoods can be calculated directly without the need to run simulations. 

To model and perform precision inference on time-varying systems is typically more difficult, especially if the underlying gravitational potential, and not only the observed stellar tracer population, is time-varying. In the absence of a steady state, it is necessary to have some other knowledge of the system. A first example is stellar streams, which form from tidally disrupted globular clusters or dwarf galaxies \citep{Johnston1995,Johnston1996,Dehnen2004}.
Crucially, the stars in a stellar stream are on similar orbits, which each coincided with a phase-space position of the parent satellite at some time in the past. Thanks to these strong constraints, they can provide precise estimates of the host galaxy's gravitational potential \citep{Johnston1999,Law2009,Law2010,Koposov2010}. A second example is the recently discovered phase-space spiral in the Milky Way disk \citep{Antoja2018}, which is a spiral-shaped density perturbation in the plane of position and velocity normal to the disk plane. Its shape is mainly determined by the disk's self-gravity, and can be used to precisely measure the vertical gravitational potential, under the approximation of a quasi steady state \citep{Widmark2021spiralI,Widmark2021spiralII,Antoja2023}. A third example, also pertaining to the Milky Way disk, is a recent work by \cite{Khalil2024}, who studied the bulk velocity field parallel to the disk plane, in order to fit a time-varying, non-axisymmetric gravitational potential with a bar and spiral arms. 
They successfully reproduce many of the observed in-plane bulk velocity features as well as the locations of the Carina-Sagittarius and Perseus arms. A fourth example is the Milky Way halo's response to being perturbed by the Large Magellanic Cloud, which can be informative of the phase-space distribution of the Milky Way dark matter halo \citep{Petersen2021,Rozier2022}.

Not only can time-varying dynamical systems provide competitive constraints, sometimes they can reveal information that would not be accessible in the corresponding steady state scenario. For example, some dark matter particle candidates have unique dynamical properties that manifest themselves on galactic scales, such as self-interacting dark matter \citep{Adhikari2022}, fuzzy dark matter \citep{Hui2017} or superfluid dark matter \citep{Berezhiani2015}. The strength and granularity of the wake in the Milky Way's dark matter halo, in response to the Large Magellanic Cloud, could provide a way to differentiate cold dark matter and fuzzy dark matter \citep{Foote2023}. A perturbed dwarf spheroidal galaxy whose mass content is dominated by a fuzzy dark matter soliton can develop a long-standing breathing mode, which does not dissipate as in the cold dark matter case \citep{Widmark2024fdm}. 

Satellite galaxies are one example where time-dependent features have not yet been exploited for dynamical mass measurements.  
These systems are particularly interesting because they are powerful probes of the dark sector \citep{Battaglia2022}, largely because they are composed of such a low fraction of baryonic matter. They are also simple systems in many regards: they have short dynamical time scales and are typically quiescent, with low amounts of gas and star formation. Constraints on dark matter models come from analyzing their internal mass density distribution (e.g. the long-standing cusp-core problem; \citealt{Popolo2021}) and population statistics \citep[e.g.][]{Garrison-Kimmel2014,Sawala2016,Fattahi2020}, as well as various indirect dark matter detection searches \citep{Strigari2018}. In the context of time-varying dynamics as a probe of the dark sector, perturbed dwarf galaxies are particularly promising.
There is a wealth of current and near future observations pertaining to the dynamics of dwarf galaxies, such as the astrometric \emph{Gaia} mission \citep{Gaia2016mission}, integral-field spectroscopic surveys like MUSE \citep{bacon2010muse} and SDSS-V's Local Volume Mapper \citep{SDSS-V_panoptic_spectro}, the upcoming Nancy Gracy Roman Space Telescope \citep{NancyGraceRoman}, et cetera. For these reasons, it is timely to develop novel methods to analyze their matter density distributions, orbits, and dynamical processes.

In this work, we consider the scenario where a dwarf galaxy satellite has been strongly perturbed during a recent pericenter passage of its host galaxy. We aim to extract information about the satellite's orbit and mass density profile by modeling its time-varying dynamics, using observations of its line-of-sight velocity field and angle of sky-projected semi-major axis, but without proper motion information. This is a challenging inference problem---crucially, we cannot directly formulate or compute a data likelihood as a function of its initial conditions. Furthermore, the model space is large, high-dimensional, and strongly degenerate.

In order to overcome these inference challenges, we employ implicit likelihood inference \citep[also known as simulation based inference, likelihood-free inference, or approximate Bayesian computation;][]{Marin2011,Cranmer2020,Ho2024}. In this framework, we circumvent the need to compute the likelihood directly. Instead, we use forward simulations in order to learn the mapping from input parameters to observables, and emulate the likelihood with neural density estimation. This is a recent development of statistical inference which, in physics and astronomy, has been applied mainly to the field of cosmology, for example in weak lensing \citep{Jeffrey2021}, galaxy clustering \citep{Makinen2022,Hahn2023}, galaxy cluster mass measurements \citep{Ho2022,Andres2022}, stellar streams \citep{Hermans2021,Alvey2023}, exoplanets \citep{Rogers2023}, and gravitational waves \citep{Dax2021}. Here we apply this framework, to our knowledge for the first time, to the internal dynamics of a tidally perturbed galaxy. 

As a testbed for our novel method, we consider the dynamics of NGC205 (also known as M110), which is a dwarf elliptical close to the Andromeda Galaxy (M31). NGC205 bears signs of a recent tidal perturbation: it exhibits S-like shapes in its phase-space distribution, both in its elongated surface brightness profile \citep{Choi2002} and its line-of-sight velocity along its semi-major axis \citep{Geha2006}. It is highly probable that these phase-space features originate from a tidal interaction with M31. Constraining the orbit and properties of NGC205 is particularly interesting, since it has been speculated to help generate M31's vast thin co-rotating plane of satellite galaxies \cite{Angus2016}; also, understanding its matter density distribution is useful for learning about the potential existence of its intermediate mass black hole \citep{Nguyen2019}. A method tailored to NGC205 would also be applicable to other dwarf galaxies with similar features. An example is NGC770, a low-luminosity dwarf elliptical close to spiral galaxy NGC772
, which has a similar S-shaped velocity profile \citep{Geha2005}.

This paper is organized as follows. In Section~\ref{sec:NGC205}, we discuss previous observations and modeling of NGC205. We present our simulation method and model of inference in Section~\ref{sec:methods}, followed by a fiducial example simulation in Section~\ref{sec:fiducial_sim}. We test our inference method on mock data in Section~\ref{sec:test_inference}, and the actual NGC205 observations in Section~\ref{sec:real_inference}. In Sections~\ref{sec:discussion} and \ref{sec:conclusion}, we discuss and conclude.

\section{Prior work on NGC205}
\label{sec:NGC205}

\subsection{Observations}

NGC205 was first discovered by Charles Messier in 1773 \citep{Jones2008}, who labelled it M110. Soon thereafter, it was also independently discovered by Caroline Herschel \citep{Herschel1785}. Its designation NGC205 comes from the New General Catalogue, first compiled in 1888 by John Louis Emil Dreyer \citep{Dreyer1888}.

\begin{figure}
    \centering
    \includegraphics[width=1.\columnwidth]{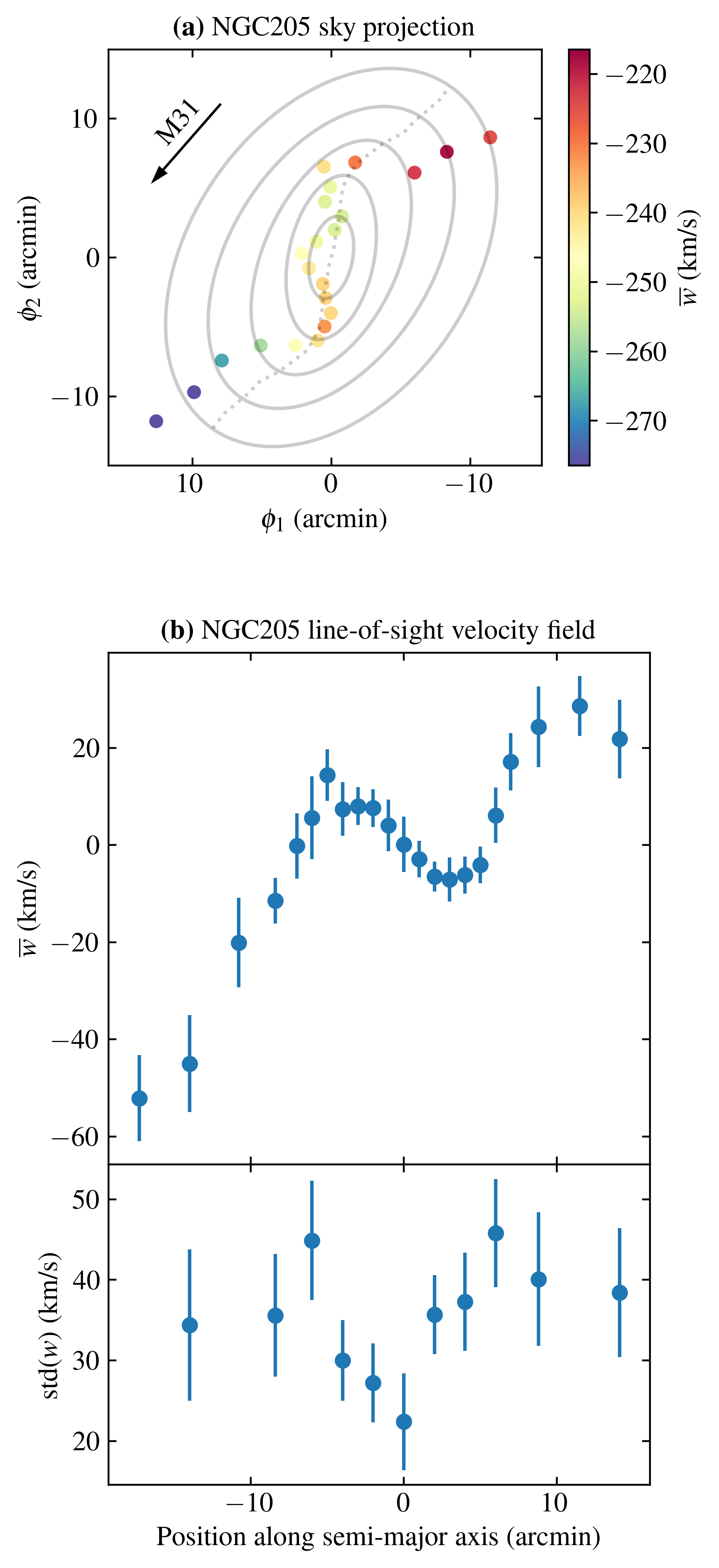}
    \caption{Observations of NGC205 from \cite{Choi2002} and \cite{Geha2006}. The top panel (a) shows the isophotes, highlighting the S-like shape seen in its sky-projected surface brightness profile. The colored dots show measurements of the mean line-of-sight velocity field. The arrow points in the direction of M31, which is at an angular distance of 54' from NGC205. The bottom panel (b) shows the mean velocity measurements as a function of spatial position along the semi-major axis, as well as the standard deviation. The standard deviation measurements have a coarser binning, to ensure a minimum of 50 individual stars per bin. See Section~\ref{sec:coord_syst} for coordinate system definitions.}
    \label{fig:real_NGC205}
\end{figure}

\cite{Choi2002} observed the photometric surface brightness of NGC205 in $B$- and $I$-bands. They fitted exponential profiles to surface brightness as a function of angular radius, and found that two different exponential scale lengths, $150\arcsec$ and $170\arcsec$, fit well over the ranges $[75\arcsec,250\arcsec]$ and $[150\arcsec,250\arcsec]$, respectively. Furthermore, they noted its S-shaped profile, describing it as a ``pronounced isophote twisting''. Assuming elliptically shaped isophotes, they fitted the isophotal orientation angle and ellipticity as a function of angular radius. Five such ellipses are visible in the top panel of Figure~\ref{fig:real_NGC205}. In \cite{Choi2002}, they stated that the isophote twisting was ``strongly suggestive of tidal interaction and probable stripping by M31''.

\cite{Geha2006} made observations of NGC205 using Keck/DEIMOS multislit spectroscopy, yielding line-of-sight velocity measurements for 725 red giant branch stars with a precision of $11.5~\kmsec$. Their chosen targets roughly lined up along the semi-major axis, following the shape of the twisting isophotes, as shown in the top panel of Figure~\ref{fig:real_NGC205}. The mean and standard deviation of the velocity field, as a function of position along the semi-major axis, are visible in the bottom panel. Intriguingly, its mean velocity profile has a clear S-like shape. It is noted by \cite{Geha2006} that the radius of the isophotal twisting matches the turn radius of the mean line-of-sight velocity field, at roughly $4.5\arcmin$. This could suggest that the two phase-space features are connected, perhaps sourced by the same tidal perturbation.

In principle, it it possible to reconcile a steady state with an S-shaped velocity profile if the inner and outer regions of the satellite are counter-rotating, although such a configuration seems highly contrived. However, a steady state is not at the same time consistent with the twisting of isophotes. As such, it would be more convincing if these phase-space features could be explained, at least in part, by a tidal interaction with M31.

In terms of constraints on the present day systemic position and velocity vector of NGC205, we use the following values, as in \cite{Howley2008}. The distance to NGC205 is $824\pm27~\kpc$, while M31 is at $785\pm25~\kpc$, giving a relative distance of $39\pm37~\kpc$ \citep{McConnachie2004}. More recent work indicate that this distance uncertainty could be underestimated and closer to $70~\kpc$ \citep{deGrijs2014}; regardless, we use the former estimate as it does not significantly alter our end results. The systemic line-of-sight velocities are $-264\pm1~\kmsec$ \citep{Geha2006} and $-300\pm4~\kmsec$ \citep{Vaucouleurs1991}, giving a relative velocity of $54\pm5~\kmsec$. While proper motions have been measured for some M31 satellites, typically with a precision of $\sim 50~\kmsec$ in the transverse velocities \citep{Sohn2020}, this has not been done for NGC205. This presents a serious difficulty in modeling NGC205, because the missing proper motion information and the poor estimate for the spatial distance gives a very large space of possible orbits.

\subsection{Modeling}

The main precursor to this work is from \cite{Howley2008}, who modeled the orbit of NGC205 with a simulation based, iterative approach. They used a genetic algorithm to fit N-body simulations to its surface brightness and velocity observations. They simulated the stellar tracer population as massless test-particles, using static models for the gravitational potentials of both M31 and NGC205. For the distribution of test-particles, they considered three different models: a cold rotating disk, a warm rotating disk, and a non-rotating pressure-supported spheroid.

Their three models could qualitatively reproduce different observed features, but not all with any single model. The rotating disk models were tailored to fit the inner velocity slope and could get an overturned outer slope through a significant tidal perturbation from M31. However, those models had velocity dispersions that were too low. Conversely, the best fit pressure-supported spheroid reproduced the velocity dispersion, but not the S-like velocity profile.

\cite{Howley2008} found that the best matching orbits have high velocities (300--500~$\kmsec$), thus placing the satellite on a very eccentric orbit, likely on its first passage of M31. They concluded that even if NGC205 starts off with a strong internal rotation which matches the inner slope of its velocity curve, a close passage with M31 is necessary in order to reverse the velocity slope in its outer region. They inferred NGC205 to be at a spatial depth of $11\pm9~\kpc$ behind M31 at present day. However, in their search they only considered a depth range of 2--76~kpc, thus completely excluding the possibility that NGC205 could be slightly in front of M31.

\newpage

\section{Methods}
\label{sec:methods}
\subsection{Coordinate systems}
\label{sec:coord_syst}

In this work, we mainly use a Cartesian coordinate system written $\boldsymbol{x} = \{x,y,z\}$, which has its origin in the host galaxy's center. Its directions are aligned with the axis of observation, such that $x$ and $y$ are along the directions of right ascension (RA) and declination (DEC), respectively, while $z$ is along the line-of-sight and increases with greater distance from the Milky Way. \update{Velocities along the same Cartesian axes are written $\boldsymbol{u} = \{u,v,w\}$. We also define two sets of sky angles which are both centered on the satellite dwarf galaxy: firstly, $\phi_1$ and $\phi_2$ are parallel with RA and DEC, respectively; secondly, $\xi_1$ and $\xi_2$ are aligned with the satellite's semi-major and semi-minor axis (further discussed in Section~\ref{sec:compressor}).}

As a tool to better understand the internal kinematics of our \update{fiducial simulation}, we also define a Cartesian orbital plane coordinate system, written $\boldsymbol{x'} = \{x',y',z'\}$. This system is also centered on the host galaxy, but $x'$ points towards the satellite's orbital pericenter, while $y'$ is pointing along the orbital trajectory at pericenter. \update{Strictly speaking, the gravitational potential of the host galaxy is not spherically symmetric, such that an orbital plane is not generally well defined. However, for our fiducial simulation, the host galaxy's disk normal is very close to parallel with the orbital plane; as a result, the satellite deviates at most a few 100~pc from the orbital plane during its passage.}

\subsection{Host and satellite modeling}
\label{sec:sat_modeling}

\update{
The host galaxy, M31, is modeled as a static gravitational potential. We use a model based on results by \cite{Zhang2024}, consisting of a Hernquist bulge, a Miyamoto-Nagai disk, and an NFW halo; see Appendix~\ref{app:M31_pot} for details about the parameter values of these components. For the inclination of M31's disk plane, we follow \cite{Howley2008}. The inclination can be found by starting from an edge-on disk, rotating it $77.5^\circ$ around the $x$-axis (axis of increasing RA), and then by $37.7^\circ$ around the $z$-axis (line-of-sight axis).
}

In our model of inference, we assume that the satellite was phase-mixed and in a steady state before its most recent pericenter passage. This is well motivated for NGC205, given that its relaxation time is significantly shorter than its orbital time. We further assume that its phase-space distribution before its most recent pericenter passage was spherically symmetric and isotropic.

We model the total mass density as a \update{heavily truncated} generalized NFW profile \citep{Navarro1996}, with the functional form
\begin{equation}\label{eq:NGC205_rho}
    \rho(r) = \frac{\rho_0 \, (r / r_h)^{-\gamma} \,
    (1 + r / r_h)^{\gamma-3}}
    {1 + (r / r_\text{tidal})^{4} }.
\end{equation}
In our inference model, the mass density $\rho_0$ and inner slope $\gamma$ are free to vary. Conversely, we fix the scale length to $r_h=4~\kpc$ and the tidal radius to $r_\text{tidal}=5~\kpc$. Varying these parameters within reasonable values has a negligible effect on the simulation results, since we are sensitive to the dynamics in the satellite's inner regions. \update{As such, only $\rho_0$ and $\gamma$ have a significant effect on the dynamics of a stellar tracer population, while for example the outer slope and truncation radius do not.} In principle, a more extended dark halo would alter the satellite's orbit somewhat, but such effects are anyway subdominant to uncertainties associated with the gravitational potential of M31.

In Table~\ref{tab:parameters}, we list the simulation input parameters that are free to vary in our model of inference. The matter density of the satellite galaxy is parametrized by a density parameter ($\rho_{1\kpc}$) and an inner density slope ($\gamma$). We use the density at a radius of one kilo-parsec, rather than $\rho_0$ in Eq.~\eqref{eq:NGC205_rho}, in order to avoid the otherwise strong degeneracy with $\gamma$. In our method of inference, we further reparametrize the density in logarithmic form, like $\log_\mathrm{10}(\rho_{1\kpc} \, \Msun^{-1}\pc^3)$. The orbit of the satellite is parametrized with the four present-day phase space coordinates $\bm{\tilde{u}} = \{\tilde{u},\tilde{v},\tilde{w}\}$ and $\tilde{z}$. The two remaining spatial coordinates ($\tilde{x}$ and $\tilde{y}$) are fully determined by the satellite's sky angles.

The prior probability distribution over these parameters, written $\Pr(\simp)$, is proportional to
\begin{equation}\label{eq:prior}
\begin{split}
    \Pr(\simp) & \propto H\{\log_\mathrm{10}(\rho_{1\kpc} \, \Msun^{-1}\pc^3) \in [-3,0]\} \\
    & \times H\{\gamma \in [0,2] \} \\
    & \times \mathcal{G}(\tilde{u},\, 300~\kmsec) \times
    \mathcal{G}(\tilde{v},\, 300~\kmsec) \\
    & \times \mathcal{G}(\tilde{w}-54~\kmsec, \, 5~\kmsec) \\
    & \times \mathcal{G}(\tilde{z}-39~\kpc, \, 37~\kpc),
\end{split}
\end{equation}
where $H\{...\}$ is a step function which is one when its argument condition is fulfilled and otherwise zero, and $\mathcal{G}(x, \sigma)$ is a Gaussian distribution with a standard deviation of $\sigma$. Only two quantities of $\simp$ are directly constrained through observations: $\tilde{w}$ is very well determined by measurements, while the constraint on $\tilde{z}$ is rather weak. \update{The transverse velocities ($\tilde{u}$, $\tilde{v}$) would be constrained by future proper motion measurements.}


\begin{table*}[ht]
    \centering
    \caption{
    Important quantities in this work.
    }
	\label{tab:parameters}
    \begin{tabular}{| l | l |}
            \hline
            $\simp$ & Simulation input parameters \\
		\hline
            $\rho_{1\kpc}$ & Matter density at $r=1~\kpc$ ($\Msunppcc$) \\
            $\gamma$ & Matter density inner slope \\
            $\bm{\tilde{u}} = \{\tilde{u},\tilde{v},\tilde{w}\}$ & 3d velocity of the satellite (km/s) \\
            $\tilde{z}$ & Position along the line-of-sight (kpc) \\
            \hline
            \hline
		$\data$ & Simulation data \\
		\hline
            $\bm{x} = \{x,y,z\}$ & 3d position (pc) \\
            $\bm{u} = \{u,v,w\}$ & 3d velocity (km/s) \\
            \hline
            \hline
		$\cdata$ & Compressed data \\
            \hline
            $\Xi$ & Angle of sky-projected semi-major axis (rad) \\
            $\xi_\mathrm{turn}$ & Turnaround point of S-shaped $\bar{w}$ curve (arcmin) \\
            $a_1$ & Value of $\bar{w}$ at $\xi_\mathrm{turn}$ ($\kmsec$) \\
            $a_2$ & Value of $\bar{w}$ at $2\xi_\mathrm{turn}$ ($\kmsec$) \\
            $s_\perp$ & Slope of $\bar{w}$ curve perpendicular to semi-major axis ($\kmsec \text{arcmin}^{-1}$) \\
            $\text{std}(w)_0$ & Standard deviation of $w$ within 3' of center ($\kmsec$) \\
            \hline
    \end{tabular}
\end{table*}

\subsection{N-body simulations}
\label{sec:N-body}

In our N-body simulations, the initial spatial distribution of particles is proportional to the mass density profile of Eq.~\eqref{eq:NGC205_rho}. The velocities are initialized isotropically using Eddington inversion \citep{Eddington1916,Lacroix2018}. We confirm that our simulations are indeed initialized in a steady state by running a test simulation which evolves without any external force; this is shown in Appendix~\ref{app:eq_test}.

In order to evolve our N-body system, we use \texttt{REBOUND} \citep{rein2012rebound}. The gravitational potential of the N-body system is solved for with a tree algorithm, which speeds up the computation. We use a simulation time-step of $2~\Myr$, and a softening length of 60~pc (a reasonable value given the mean nearest neighbor distance, see e.g. \citealt{Athanassoula2000}).

We use a total of 120,000 massive particles. We also add 40,000 massless particles at small radii, distributed proportional to Eq.~\ref{eq:NGC205_rho} with parameter values $\gamma=1$, $r_h=1~\kpc$, and $r_\text{tidal} = 2~\kpc$. They are added to ensure that there are enough particles at low radii to construct the tracer particle population. This construction is described in detail in Section~\ref{sec:find_tracer} below.

\subsubsection{Back propagation}
\label{sec:back_prop}

The phase-space coordinates in $\simp$ ($\tilde{u}$, $\tilde{v}$, $\tilde{w}$, $\tilde{z}$) correspond to the satellite's systemic position at $t=0$. In order to model the satellite's approach to this point, we first back propagate its orbit. This back propagation is performed for a single particle in the static external potential of the host galaxy, going back 500~Myr. From that starting position, the satellite is then evolved from its initial steady as a full N-body simulation, from $t=-500~\Myr$ to $t=0~\Myr$.

Since the satellite galaxy is subject to tidal distortion and stripping, its orbit is not identical to that of a single test particle. This gives rise to a small mismatch between the phase-space parameters and the simulated satellite's center at $t=0$. We test the magnitude of this mismatch with our fiducial simulation (which is described in detail in Section~\ref{sec:fiducial_sim} below).

For the fiducial simulation, the spatial displacement in $\tilde{z}$ is 36~pc, and at most 50~pc in $\tilde{x}$ and $\tilde{y}$. For the velocities, $\tilde{w}$ is most affected with an offset of $-2.8~\kmsec$, while the transverse velocities $\tilde{w}$ are shifted by sub-$\kmsec$ values. Shifts of this magnitude do not have a significant effect on our inference results, and are definitely subdominant with respect to potential biases associated with the gravitational potential of M31 (see Appendix~\ref{app:M31_pot}).

\subsubsection{Tracer population and observational errors}
\label{sec:find_tracer}

The massive particles of the simulation represent the total mass density of the satellite, which includes both baryonic and dark matter. However, what we can observe is a stellar tracer population. In our simulations, the tracer population is selected as a sub-population by fitting and applying an inclusion probability function, written $I(E_i)$.

The function $I(E_i)$ is the probability for massive and massless particles to be included in the tracer population. It is fitted to the simulation's final state, such that the tracer population matches NGC205's sky-projected surface brightness profile. The inclusion function depends only on initial energy ($E_i$), which ensures that the tracer population is initialized in a steady state. This circumvents the need to run a completely separate population of tracer particles, and we can reproduce the surface density radial profile without the need to include additional parameters in our inference framework.

In more precise terms, $I(E_i)$ is a mixture model of three Gaussians, which we fitted to reproduce a target surface density that decays exponentially with a 170'' scale length (as observed for NGC205, see Section~\ref{sec:NGC205}). This is done by constructing a histogram of sky-projected angular radii in 20 bins evenly spaced between 0'--20', and performing a least-square fit compared to the target histogram. In this manner, $I(E_i)$ is only fit in terms of its radial dependence, but not its elongation or other spatial asymmetries.

After this fit is performed, we further limit the stellar tracer population in two steps, in order to create a realistic total number count of observations. Firstly, we decrease it to a total of 8,000 particles, a subset representing photometric observations. This thinning ensures that different simulations end up with the same data count and statistical power, in a manner that does not depend on how strongly the simulated satellite was perturbed. Secondly, we set an upper limit of 28 particles per $2\arcmin \times 2\arcmin$ angular area bin. In each area bin, these particles are randomly selected from the tracer population, and represent the available velocity information. This restriction results in an inner region, within roughly 10', where the velocity field is uniformly sampled, and an outer region where the velocity observation density decreases exponentially with radius. This mimics a real data taking scenario, where the number of velocity measurements is typically not proportional to the surface brightness profile (in such a case, the measurements would be much more concentrated towards the satellite's center).

In our simulations, we end up with a total number of roughly 3,200 velocity measurements. To each star, we apply a velocity error of $11.5~\kmsec$, corresponding to the Keck observation uncertainties in \cite{Geha2006}.

\subsection{Inference model}
\label{sec:inference_model}


\update{Extracting information from NGC205's time-varying dynamical features is a difficult inference problem.}
Firstly, without simulations we cannot formulate or write down a precise connection between observable features and the initial conditions we are interested in inferring. Secondly, the model-to-data connection is highly degenerate, so we cannot isolate the effects of the respective initial condition parameters. Thirdly, the data itself is high-dimensional and complicated, so how to formulate a goodness-of-fit is a non-trivial issue. Fourthly, key observables are missing, for example proper motion data that would constrain the satellite's orbit. As a result, the full model space is very large compared to the target space of simulations that actually fit the data. To search the model space manually can give some insights, but ultimately we want to automate this search and to perform precise and rigorous inference.

In order to surmount these challenges, we employ a Bayesian inference framework of forward simulation with likelihood emulators 
\citep{Marin2011,Cranmer2020}.
The defining characteristic of this type of inference is that it circumvents any explicit computation of the likelihood probability. Instead, the likelihood is emulated, in our case by training a neural density estimator on the results of forward simulations, thus learning the mapping from initial parameters to observable quantities.
It is not tractable to emulate the likelihood for the full data set that is produced by simulations. Instead, we use a compressor to reduce the data to a handful of summary statistics that are deemed informative. In this work, we run N-body simulations with roughly $10^5$ particles, and select a sub-sample of a few thousand stars to represent realistic observations. This data ($\data$) is then compressed to only six parameters (the reduced data, $\cdata$).

In our case, it is too expensive to explore the full $\simp$ parameter space, as set by the prior $\Pr(\simp)$. To overcome this, the inference framework works iteratively. 
\update{In each step of the inference loop, we run a new batch of simulations. With every new batch, the likelihood emulator refines its 
understanding of the mapping $\simp \rightarrow \cdata$. We then sample the posterior probability (given by the prior and emulated likelihood), in order to draw a new batch of $\simp$ values to simulate. With each step of the inference loop, we are honing in on the region of $\simp$-space which gives rise to $\cdata$ consistent with the given target, $\cdata_\mathrm{target}$. Thus we are, to increasing degree, running simulations in the region of interest. Eventually, the posterior distribution converges and becomes stable between successive batches of simulations. The various aspects of the inference model are described in detail below.}

\subsubsection{Data compressor}
\label{sec:compressor}

\update{The compressor, written $C:\data\rightarrow\cdata$, takes the N-body simulation data ($\data$), which has a dimensionality of $\mathcal{O}(10^4)$, and reduces it to a compressed data vector with only six parameters ($\cdata$, listed in Table~\ref{tab:parameters}).} The reduced data $\cdata$ encapsulates the angle of the satellite's semi-major axis ($\Xi$), four parameters which describe the mean line-of-sight velocity field ($\xi_\mathrm{turn}$, $a_1$, $a_2$, $s_\perp$), and the central velocity standard deviation ($\text{std}(w)_0$). These specific parameters were chosen by hand, deemed to be both easily interpretable and informative. In contrast, it would be possible, for example, to model the compressor with a neural network, and train that to find a maximally informative compressed data vector. Regardless, we keep $C$ as ``hand-picked'' in order to retain interpretability and control. This way we ensure that the inference is based on physically realistic features, rather than some artifact, texture, or microscopic features that could depend on simulation details.

The angle of the satellite's sky-projected semi-major axis is written $\Xi$. It is given by the angle of covariance of the sky-projected tracer population (i.e. the photometric subset of 8,000 particles). The central velocity standard deviation, $\text{std}(w)_0$, is given by the tracer population subset with velocity measurements within 3' of the satellite's center.

The mean line-of-sight velocity field, as parametrized by four of our $\cdata$ parameters, is given by
\begin{equation}\label{eq:2d_vel_model}
    \bar{w} (\phi_1,\phi_2) = S(\xi_1) + s_\perp \xi_2,
\end{equation}
where
\begin{equation}
    \begin{split}
        \xi_1 & = \cos(\Xi)\phi_1 + \sin(\Xi)\phi_2, \\
        \xi_2 & = -\sin(\Xi)\phi_1 + \cos(\Xi)\phi_2,
    \end{split}
\end{equation}
are sky angle coordinates that are parallel and perpendicular to the semi-major axis. The function $S(\xi_1)$ is a spline which models the S-like profile of the mean velocity field, equal to
\begin{equation}\label{eq:vel_spline}
    \begin{split}
    & S(\xi_1) = \\
    & \begin{cases}
        \dfrac{2 a_1}{\xi_\mathrm{turn}} \bigg( \xi_1-\dfrac{\xi_1^2}{2\xi_\mathrm{turn}} \bigg)
        & \text{for } \xi_1 \in [0', \xi_\mathrm{turn}], \\
        a_1 + \dfrac{a_2-a_1}{\xi_\mathrm{turn}^2} ( \xi_1-\xi_\mathrm{turn} )^2
        & \text{for } \xi_1 \in (\xi_\mathrm{turn}, 2\xi_\mathrm{turn}], \\
        a_2 + \dfrac{2(a_2-a_1)}{\xi_\mathrm{turn}} (\xi_1-2\xi_\mathrm{turn})
        & \text{for } \xi_1>2\xi_\mathrm{turn}, \\
        -S(|\xi_1|)
        & \text{for } \xi_1<0.
    \end{cases}
    \end{split}
\end{equation}
This spline is continuous in value and first derivative. It is anti-symmetric and passes through points $(0',0~\kmsec)$, $(\xi_\mathrm{turn}, a_1)$, and $(2\xi_\mathrm{turn}, a_2)$. It has a zero-valued derivative at $\xi_\mathrm{turn}$, and is linear beyond $2\xi_\mathrm{turn}$.

This function is fitted to the simulated line-of-sight velocity data. The velocity data is binned in $2\arcmin \times 2\arcmin$ area bins (after thinning the data and applying errors, as described in Section~\ref{sec:find_tracer}). We use least-square minimization, where the statistical variance of an area bin is inversely proportional to its particle number. An example of this fit is presented in Section~\ref{sec:fiducial_sim} below.

\subsubsection{Neural density estimators}
\label{sec:NDEs}

Neural density estimators are flexible models that depend on network parameters $\boldsymbol{\varepsilon}$. We use a Gaussian mixture density network, where the network inputs are $\simp$ and the network outputs are the means and covariance factors of a Gaussian mixture model. This Gaussian mixture model, written $\text{Pr}(\cdata \, | \, \simp, \boldsymbol{\varepsilon})$, emulates the conditional likelihood density $\mathcal{L}(\cdata \, | \simp)$. The network parameters $\boldsymbol{\varepsilon}$ are trained on a set of $\{ \simp,\cdata \}$ pairs.

We use an ensemble of four networks with identical architecture, which are trained and used in parallel to avoid over-fitting artifacts of a single network. We use the neural density estimators that are implemented in the DELFI framework as \texttt{MixtureDensityNetwork}, with the following network architecture values:
\begin{itemize}
    \item $\texttt{n\_components}=5$,
    \item $\texttt{n\_hidden}=[80,80]$,
    \item $\texttt{activations}=[\texttt{tf.tanh}, \texttt{tf.tanh}]$.
\end{itemize}
We refer to \cite{Alsing2019} for further details.

\subsubsection{Inference loop}
\label{sec:inference_loop}

\begin{figure*}
    \centering
    \includegraphics[width=.85\textwidth]{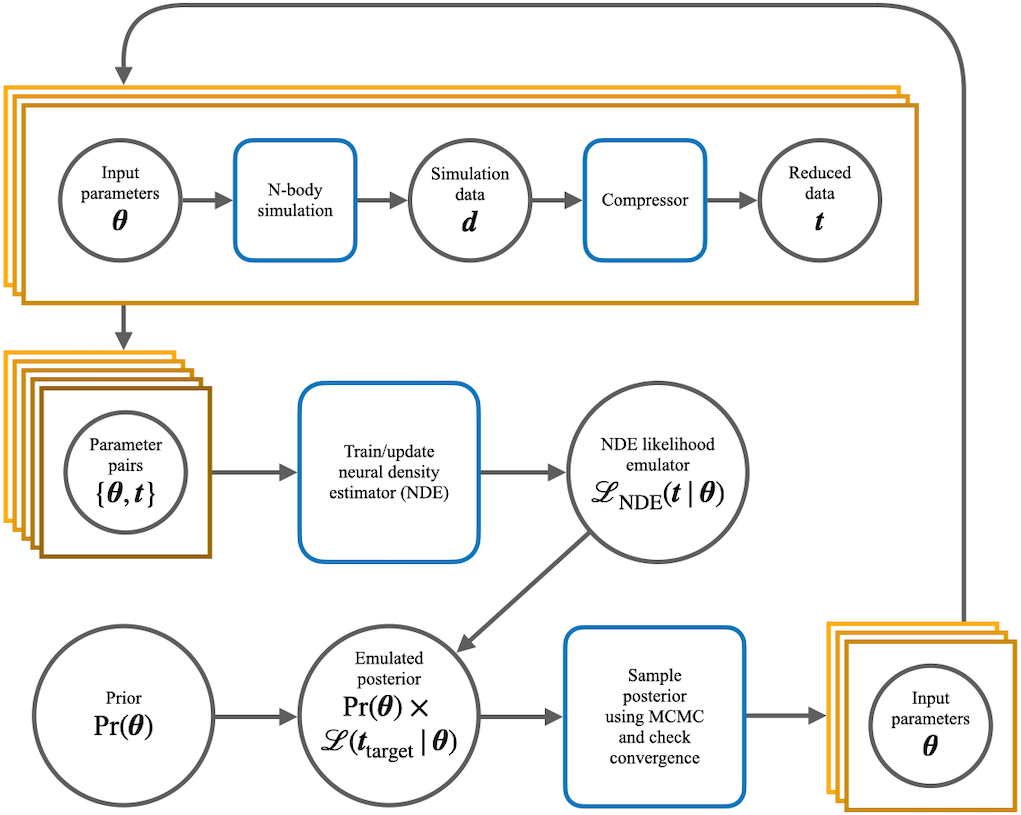}
    \caption{Flowchart representing the inference process loop. Stacked squares represent a batch with several realizations; the stack in the second row has more levels, representing many batches accumulated over all loop iterations. See the main text for further details.}
    \label{fig:flowchart}
\end{figure*}

We start the inference process by drawing a first batch of 1000 $\simp$ realizations from the prior probability distribution. We then loop through the following steps:
\begin{enumerate}
    \item Run \update{a batch} simulations and compress the data, giving a \update{new} batch of $\{\simp,\cdata\}$ pairs.
    \item Train the neural density networks to emulate the likelihood, $\mathcal{L}_\mathrm{NDE}(\cdata \, | \, \simp)$. \update{This training uses all batches of $\{\simp,\cdata\}$ pairs (i.e. from all previous iterations of the inference loop).}
    \item Check to see if the posterior probability density has converged, by producing roughly $10^5$ realizations of $\simp$ from the emulated posterior. If not yet converged, draw a new batch of 200 $\simp$ realizations from the posterior \update{(only the zeroth batch has size 1000). This new batch of $\simp$ values will be simulated in the next iteration of the inference loop.}
\end{enumerate}
This inference loop is visually represented in a flowchart in Figure~\ref{fig:flowchart}. The dominant computational cost of the whole inference process is running the N-body simulations in the loop's first step.

In the third step of our inference loop, we sample $\simp$ from the posterior density, meaning the emulated likelihood times the prior, as expressed in Eq.~\ref{eq:prior}: $\Pr(\simp\, | \cdata_\mathrm{target}) = \Pr(\simp)\mathcal{L}_\mathrm{NDE}(\cdata_\mathrm{target} \, | \, \simp)$. We use the ensemble Markov chain Monte Carlo (MCMC) implemented in the DELFI package, called \texttt{emcee\_sampler}.

\update{We test for posterior distribution convergence by comparing the posterior samples (a group of roughly $10^5$ realizations drawn by MCMC) with those of previous inference loop iterations. For our stopping criteria, we require the posterior to be stable over ten successive loop iterations.} 
This stability is achieved when the standard deviations of the respective $\simp$ parameters no longer decrease, within a one per cent margin. We do the same test for the determinant of the $\simp$ covariance matrix. We also ensure convergence by studying the posterior density distribution of the final iteration steps by eye. When these stopping criteria are fulfilled, we take the emulated posterior of the final step to be the posterior probability distribution.

\section{Fiducial example simulation}
\label{sec:fiducial_sim}
In this section, we show our fiducial example simulation. This simulation is not to an exact quantitative fit to the actual data of NGC205. Rather, it was chosen by hand in order to reproduce NGC205's features qualitatively, and to test our model of inference. For this fiducial example, we use initial parameter values:
\begin{itemize}
    \item $\rho_{1\kpc} = 10^{-1.2}~\Msunppcc \simeq 0.063 ~\Msunppcc$,
    \item $\gamma=1$,
    \item ${\bm{\tilde{u}}} = \{-340, 370, 54\}~\kmsec$,
    \item $\tilde{z}=-9~\kpc$.
\end{itemize}

The main purpose of this section is to illustrate and build intuition for the dynamical process. For this reason, the results presented in this section do not include the full treatment of data uncertainties. We do select a tracer population as described in Section~\ref{sec:find_tracer}, but after that we do not apply further data thinning or any observational errors. The full treatment of data uncertainties is applied when we test our inference model on mock data in Section~\ref{sec:test_inference} below.

\subsection{Fiducial simulation in orbital plane coordinates}

\begin{figure*}
    \centering
    \includegraphics[width=1.\textwidth]{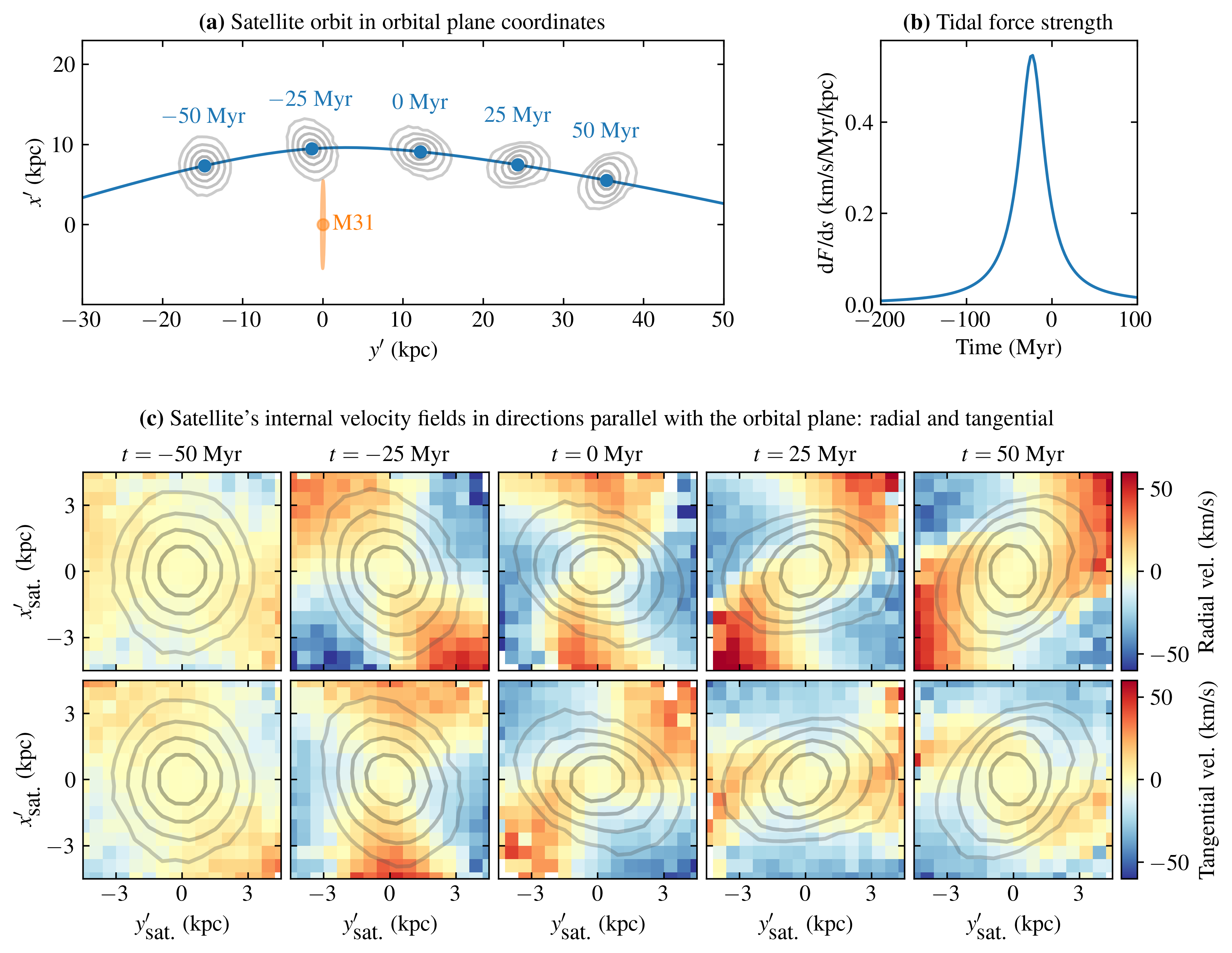}
    \caption{
    The orbit and orbital plane velocity fields of our fiducial simulation. Panel (a) shows a section of the satellite's orbit in orbital plane coordinates $x'$ and $y'$, with five highlighted snapshots in time. The grey contour lines show the tracer particle surface density, at intervals of 0.4 dex. The stellar disk of M31 is shown as an ellipse; in this projection it is close to edge on. Panel (b) shows the tidal force strength as a function of time. Finally, panel (c) contains sub-panels showing the satellite's internal velocity field for the five snapshots; the top and bottom rows of sub-panels show the radial and tangential velocity fields in the orbital plane, \update{centered on the satellite and in the satellite's rest frame. We stress that neither of these velocities are parallel with an observer's line-of-sight; see the main text for further details.} The overlaid gray contours are the same as those in panel (a).
    }
    \label{fig:sim_orbit}
\end{figure*}

In Figure~\ref{fig:sim_orbit}, we show the orbit and dynamical evolution of our fiducial simulation in orbital plane coordinates \update{(as defined in Section~\ref{sec:coord_syst})}. Hence, it does not correspond to the frame from which NGC205 is actually observed. The bottom panel shows \update{two of} the simulated satellite's internal velocity fields for five different snapshots in time, centered on $t=0~\Myr$. \update{The two velocities are radial and tangential in cylindrical coordinates (i.e. outward and rotational, respectively) and are both parallel with the orbital plane, in a coordinate system that is centered on and co-moving with the satellite.} As the satellite approaches pericenter, it becomes elongated, and develops a quadrupole in the radial velocity field. Because the tidal force is not only changing in amplitude, but also in angle, a quadrupole develops also in the rotational velocity field. The satellite galaxy's inner parts are more shielded from the external tidal impulse and have shorter dynamical time-scales, resulting in a twisting of the dipole structure in both velocity fields.

A full description of the phase-space distortion is high-dimensional and rather complex. It jointly depends on the amplitude and angle of the tidal force, as set by the orbit and external gravitational potential, as well as the satellite's self-gravity and the initial configuration of the tracer population. To complicate things further, what these phase-space features look like to an observer is highly dependent on viewing angle and timing. The line-of-sight vector needs to be sufficiently close to parallel with the orbital plane in order \update{for these features to be apparent in the line-of-sight velocity field.}

The total matter density parameters ($\rho_{1\kpc}$ and $\gamma$) determine how responsive the inner and outer parts of the satellite galaxy is to an external perturbation. Although it is difficult to visualize given the degeneracies with other parameters, we can at least assert that there is a connection between the matter density parameters and the velocity field quadrupoles and their degree of twisting. Not all choices of $\rho_{1\kpc}$ and $\gamma$ can give rise to the velocity profiles seen in Figure~\ref{fig:sim_orbit}; thus it should be possible, at least in principle, to infer those parameters.

\subsection{Fiducial simulation in the observational frame}

\begin{figure*}
    \centering
    \includegraphics[width=1\textwidth]{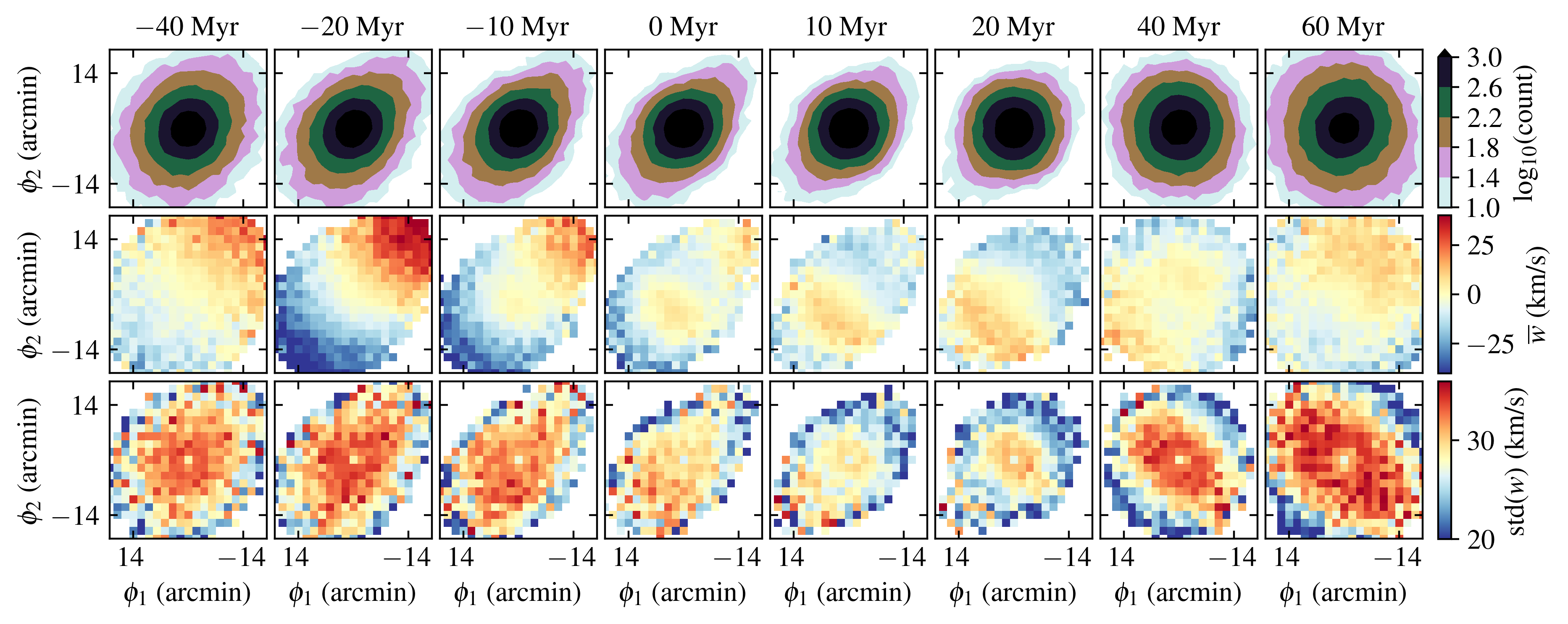}
    \caption{Observable surface density and velocity fields of the fiducial simulation. Each column shows a different snapshot in time close to the ``present moment'' of $t=0~\Myr$. The rows correspond to, from the top, tracer particle surface density, mean line-of-sight velocity, and line-of-sight velocity standard deviation. The axes are shared between all panels, and denote sky angles relative to the satellite's center.}
    \label{fig:sim_obs_proj}
\end{figure*}

In Figure~\ref{fig:sim_obs_proj}, we show the fiducial simulation from the view point of the observer (i.e. in sky-projected spatial position and line-of-sight velocity), for eight different snapshots in time. The mean velocity field, seen in the middle row, develops a dipole as the satellite approaches pericenter at $t\simeq -20~\Myr$. After that, an S-like velocity profile evolves along the semi-major axis. It starts with a small inner amplitude ($a_1$) and turn-around radius ($\xi_\mathrm{turn}$), which both grow over a time-scale of roughly 30~Myr. The surface brightness profile temporarily develops a highly elongated shape, also lasting a few 10~Myr. Lastly, the velocity dispersion field undergoes a significant change, reaching low values especially around roughly 0--20~Myr.

\begin{figure}
    \centering
    \includegraphics[width=1.\columnwidth]{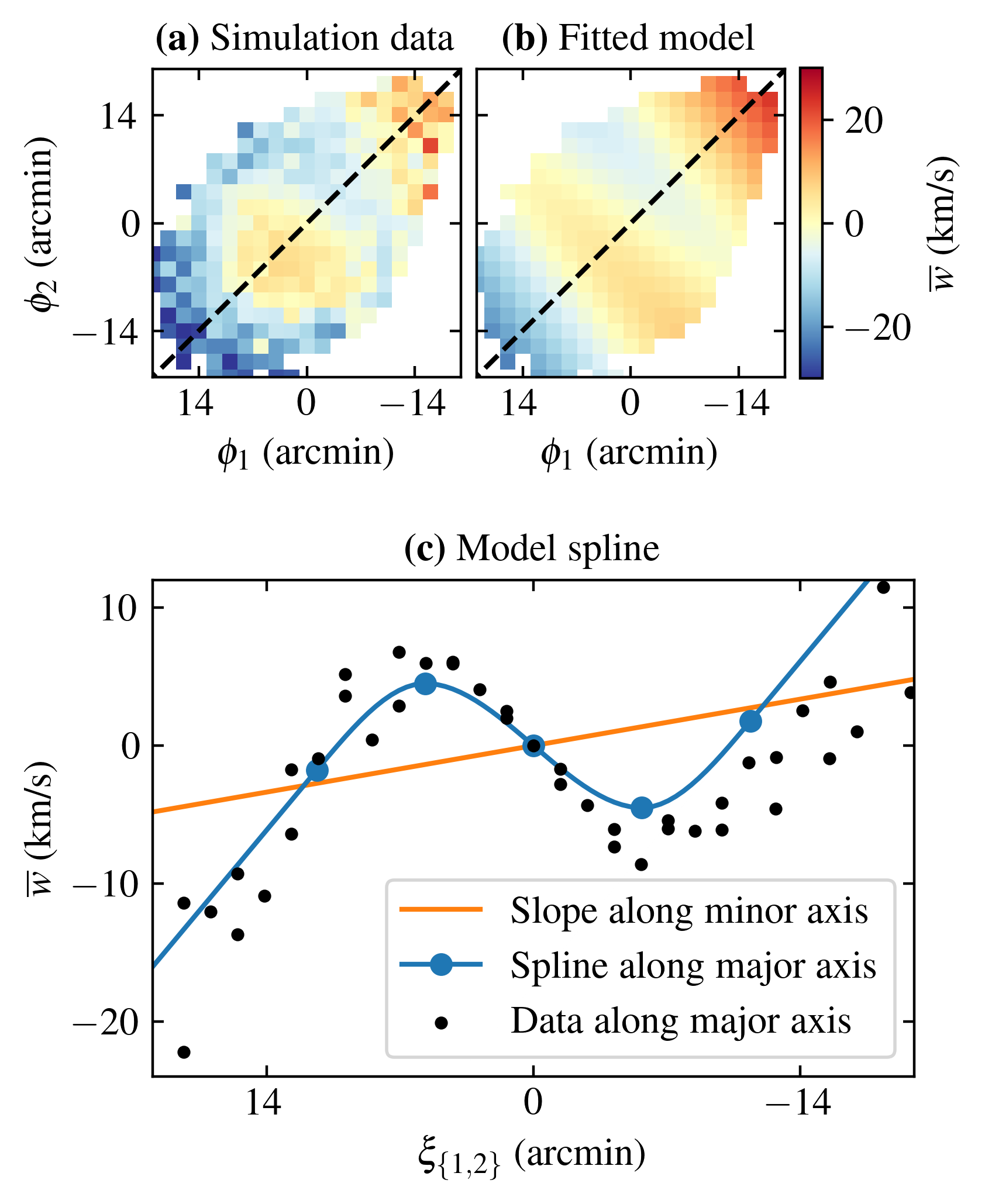}
    \caption{Mean line-of-sight velocity field and the fitted model of the compressor, for our fiducial simulation at $t=0~\Myr$. The top panels show the mean velocity field (also seen in Figure~\ref{fig:sim_obs_proj}) and fitted model. The orientation of the semi-major axis, parametrized by $\Xi$, is shown as a dashed black line. The bottom panel shows the fitted S-shaped spline along the semi-major axis, $S(\xi_1)$, and the fitted slope along the semi-minor axis, parametrized by $s_\perp$. Overlaid black points show the data along the semi-major axis, for area cells within 2.8' of the dashed black line in panel (a). The observational errors that we otherwise apply to our simulations are not used here in order to better illustrate the fit and its comparison with actual data features; see the main text for further details. 
    }
    \label{fig:spline}
\end{figure}

The S-like velocity profile is further illustrated in Figure~\ref{fig:spline}, along with the velocity field model of the compressor, as written in Eq.~\eqref{eq:2d_vel_model}. As mentioned above, we do not apply observational errors in this particular section, in order to better illustrate the phase-space features; in our full treatment of observational errors the uncertainties are roughly $6~\kmsec$ for the central area bins, while here they are around $1.8~\kmsec$. As seen in the top panels, the fitted model does not fully capture the structure of the simulated data, especially in the outer parts in the direction perpendicular to the semi-major axis. The fact that the velocity model does not reproduce all the features of the data is not in itself a problem, as long as the simulated data and target data are compressed in the same manner. The velocity model is fitted over the full field, but the statistical power comes mainly from the region within 10', which drives the fit. Because the fit is made over this larger sky area, the fitted spline has a slightly smaller amplitude compared to the data that goes precisely along the semi-major axis.

Our fiducial simulation reproduces many of NGC205's observed phase-space features. Most importantly, we clearly see an S-like mean velocity profile along the semi-major axis, qualitatively similar to that seen in NGC205. We see a good agreement in terms of $\Xi$ and $\text{std}(w)_0$.

We also reproduce some features that are not accounted for in our inference model. The ellipticity in surface brightness profile, meaning the ratio of the semi-minor and semi-major spatial axes, is similar in the outer region ($\gtrsim10'$). Although the NGC205 data is rather noisy, the velocity standard deviation field seems to be in good agreement, with a local minimum in the inner regions. We also see a hint of the same break from anti-symmetry in the mean velocity profile: comparing Figures~\ref{fig:real_NGC205} and \ref{fig:spline}, the outer slope on the right hand side is less steep than on the left hand side.


Conversely, some features are not well reproduced by our fiducial simulation. The most crucial shortcoming of our fiducial simulation is the absence of twisting isophotes. Furthermore, the surface density profile in the simulated satellite's inner region is less elliptical than in its outer region, while the opposite is true for NGC205. In terms of the S-shaped velocity profile, it is not a perfect match quantitatively: NGC205 has a shorter turnaround point ($\xi_\mathrm{turn}$) and steeper velocity slopes, both in the inner and outer regions.

\subsection{Adding internal rotation to the tracer population}
\label{sec:internal_rotation}

\begin{figure}
    \centering
    \includegraphics[width=0.78\columnwidth]{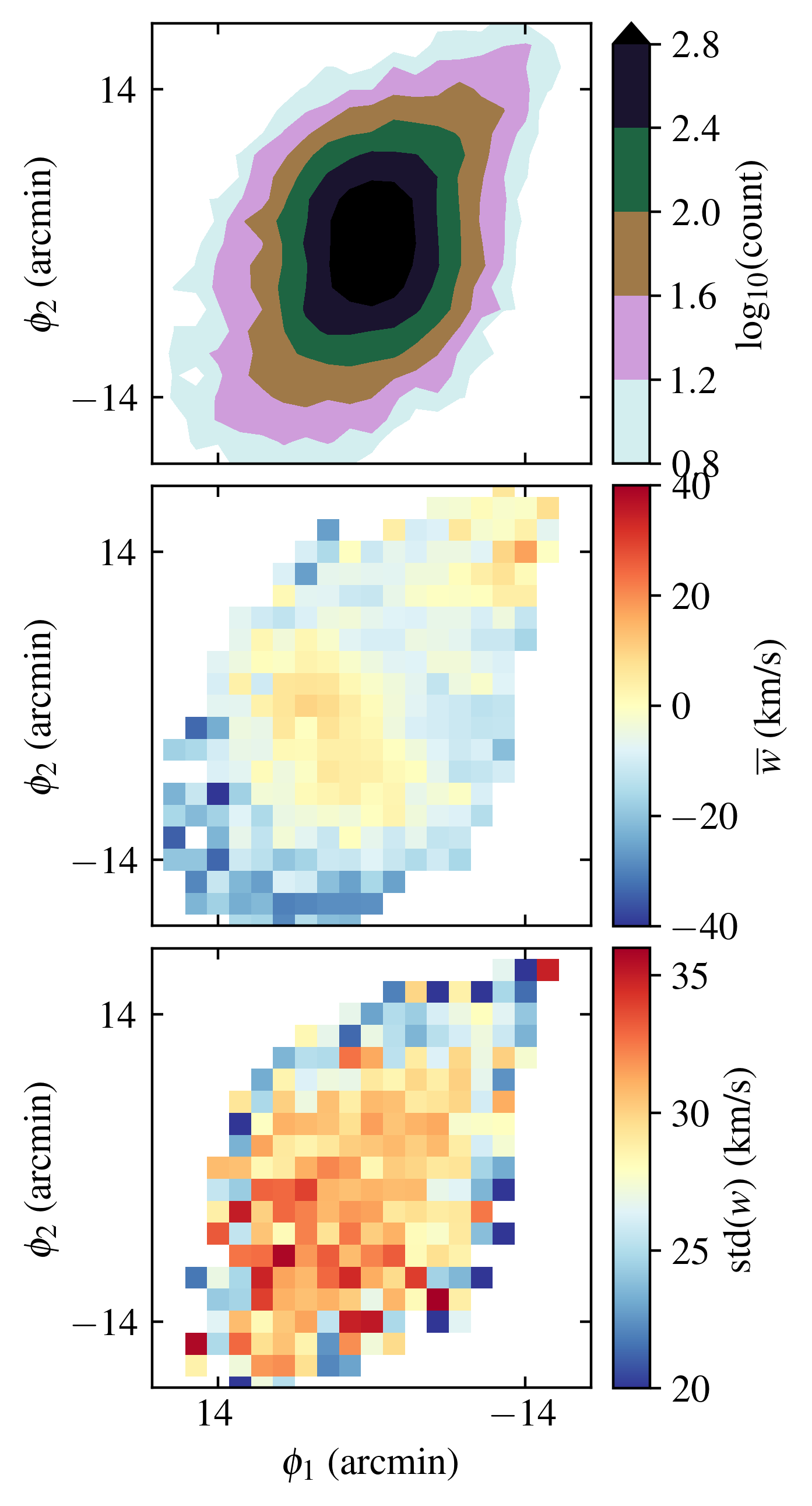}
    \caption{Observable surface density and velocity fields of our fiducial simulation at time $t=0~\Myr$, equivalent to the fourth column in Figure~\ref{fig:sim_obs_proj}, although here modified such that the tracer population is initialized with a moderate amount of internal rotation. As can be seen in the top panel, this gives rise to an isophotal twist, similar to the observed surface brightness profile of NGC205.}
    \label{fig:sim_with_rot_small}
\end{figure}

We performed some tests where we relaxed the assumptions of our initial conditions, to see if we could reproduce the observed twisting isophotes of NGC205. Specifically, we ran simulations where we added internal rotation to the tracer population of the fiducial simulation. Dwarf ellipticals like NGC205 often do have a degree of rotation; for example, see \cite{Scott2020} for population statistics of rotationally supported and dispersion supported systems.

The tracer population's internal rotation is created by making cuts in angular momentum. Using the initial state of the tracer particle distribution in the satellite rest frame, excluding the external potential, we calculate the respective particles' maximum angular momentum ($L_\mathrm{max}$) as given by their total energy (i.e. as if the particle was on a perfectly circular orbit, \update{going clockwise such that $L_\mathrm{max}$ is positive}). We then exclude tracer particles by making a hard cut in the ratio of angular momentum of some rotation axis divided by the maximum angular momentum. Because we construct the tracer population by making cuts only in angular momentum and initial energy, we ensure that the tracer population is initialized in a steady state.

In Figure~\ref{fig:sim_with_rot_small}, we show results for the fiducial simulation at $t=0~\Myr$, where we have made the cut $L_z/L_\mathrm{max}<0.1$, using a rotation axis which is parallel to the observer's line-of-sight. This cut removes 40~\% of the tracer population particles. In the top panel, we clearly see twisting isophotes qualitatively similar to those of NGC205. In the middle panel, we also see that the mean line-of-sight velocity field is pretty much preserved, although slightly lopsided. A larger figure with more time snapshots, analogous to Figure~\ref{fig:sim_obs_proj}, is shown in Appendix~\ref{app:internal_rotation}. Choosing other rotation axes can have much more dramatic effects on the velocity field, for example making the S-like shape more pronounced or giving rise to strong asymmetries across the semi-major axis.

This coarse treatment, with a hard cut in angular momentum, is not meant to be a realistic model. Neither is it fine-tuned to precisely fit the NGC205 observations. Rather, this simple experiment serves to illustrate the general point that twisting isophotes can arise if a moderate amount of rotation is included in the observable tracer's initial state. Internal rotation and fitting the isophotal twist is not currently implemented in our inference model but we plan to do so for future work; this is further discussed in Section~\ref{sec:discussion}.

\section{Test of the inference model with simulated data}
\label{sec:test_inference}

We test our inference model using the fiducial simulation, running a total of 10,800 simulations before reaching the convergence criterion. We have also checked for convergence by eye, seeing that the inferred posterior is stable over the final ten simulation batches.

\begin{figure*}
    \centering
    \includegraphics[width=0.95\textwidth]{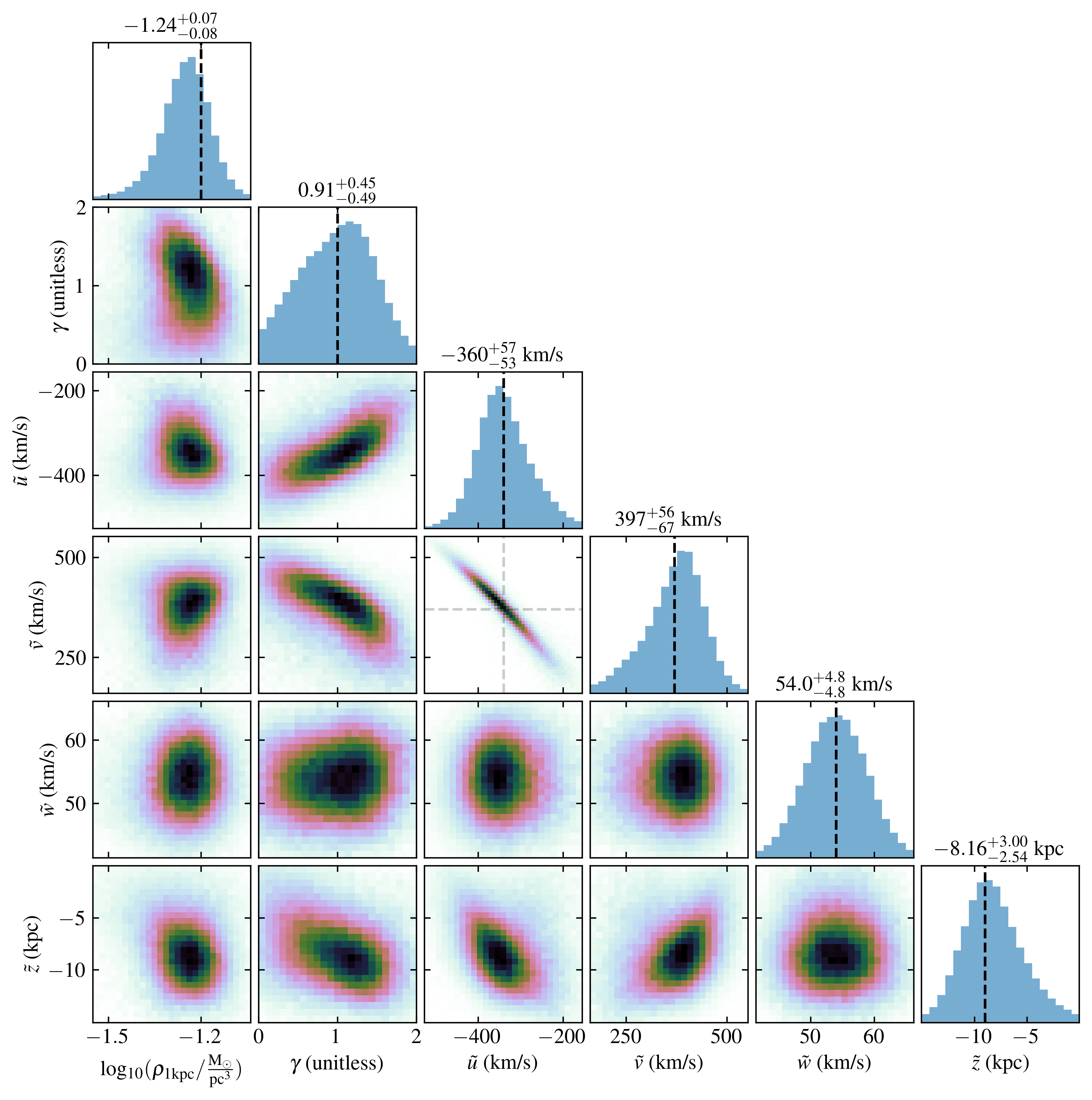}
    \caption{The inferred posterior density in 1d and 2d marginalizations, for our tests on simulated data using the fiducial simulation as target. The dashed line in the 1d histograms, as well as the $\tilde{u}$--$\tilde{v}$ 2d histogram, show the true target value. The numbers on top are the median of the marginalized posterior, plus/minus the difference with respect to the 16th and 84th percentiles. The bin sizes of the respective 2d histograms are adaptive, such that 80~\% of the posterior is contained within 200 bins.}
    \label{fig:sim_posterior}
\end{figure*}

\subsection{Posterior distributions}
We show the inferred posterior in Figure~\ref{fig:sim_posterior}. All of the simulation input parameters are retrieved well within one standard deviation of the posterior probability mode. The matter density amplitude ($\rho_{1\mathrm{kpc}}$) is inferred with an uncertainty of 0.08 in log$_{10}$-space, which corresponds to a linear relative uncertainty of 18\%. We have some constraining power on the matter density slope: for example, a cuspy profile ($\gamma=1$) is correctly preferred over a cored profile ($\gamma=0$). We can also see that $\gamma$ is quite degenerate with the transverse velocities ($\tilde{u}$ and $\tilde{v}$). Therefore, systemic proper motion measurement would be useful not only for confirming the inferred orbit, but also in providing a slightly better constraint on $\gamma$.

The inferred transverse velocities ($\tilde{u}$ and $\tilde{v}$) are highly degenerate with each other. The direction of the satellite's orbit is tightly constrained, mainly by the angle of its spatial elongation (as parametrized by $\Xi$), while its total speed is more free to vary. The individual parameters $\tilde{u}$ and $\tilde{v}$ both have standard deviations of roughly 60~km/s. If we instead consider the transverse velocity along the degenerate axis, the standard deviation is 90~km/s. The marginalized posterior for the systemic line-of-sight velocity ($\tilde{w}$) is strongly prior driven. Varying $\tilde{w}$ within its narrow prior range does not have a significant effect on our inferred results. Finally, the systemic spatial depth with respect to the host galaxy ($\tilde{z}$) is also precisely and accurately inferred.

\begin{figure}
    \centering
    \includegraphics[width=1.\columnwidth]{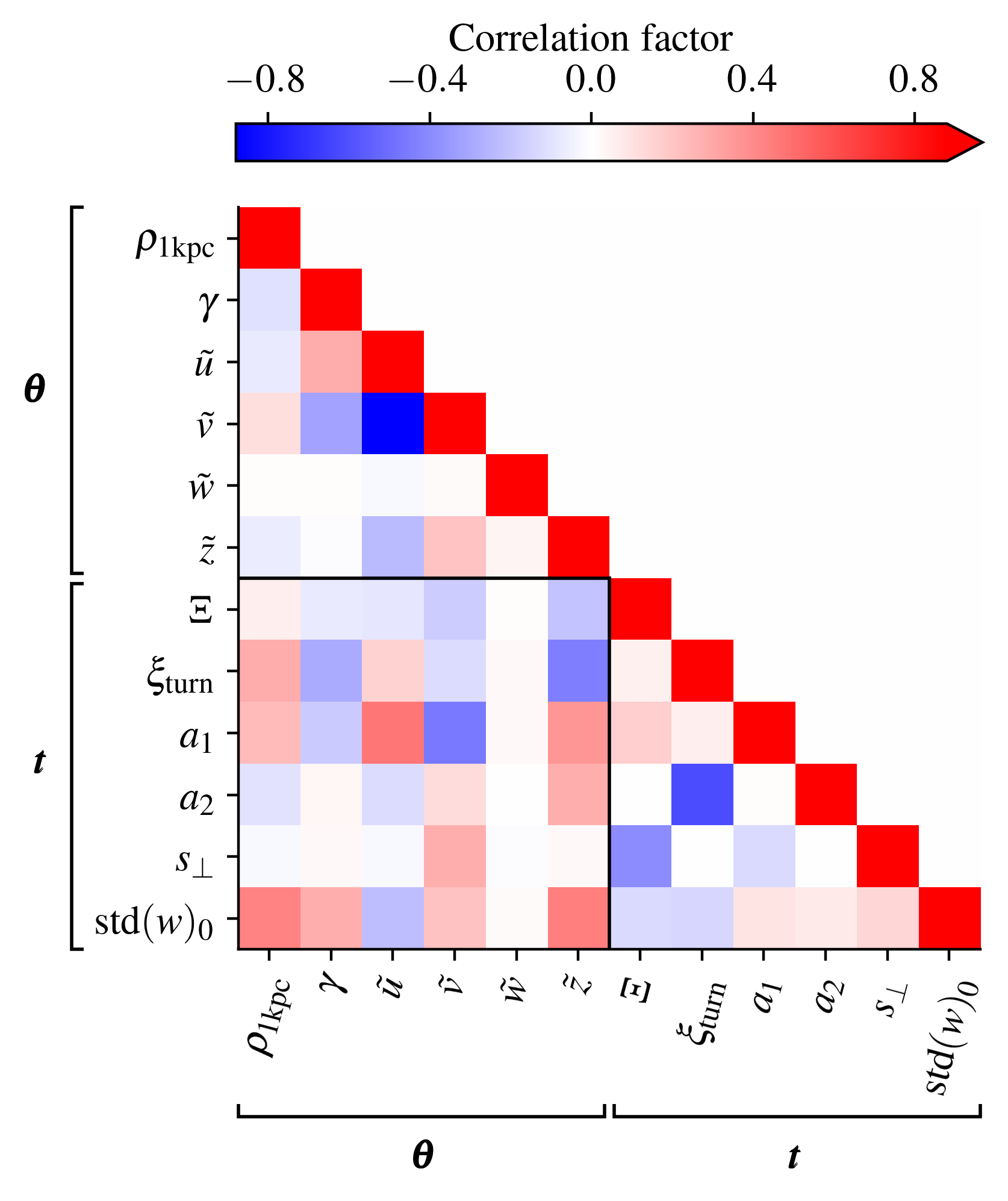}
    \caption{
    Correlations of $\simp$ and $\cdata$, for our inference on simulated data. These correlations are calculated from simulations that are sufficiently close to the mode of the inferred posterior distribution; see the main text for further details. The black lines separate $\simp$ and $\cdata$.
    }
    \label{fig:corrs}
\end{figure}

\subsection{Relating input and reduced data parameters}
Correlations between input parameters ($\simp$) and reduced data parameters ($\cdata$) are shown in Figure~\ref{fig:corrs}. In order to compute these correlations, we first calculate the covariance matrix of $\simp$ in the inferred posterior density distribution. We then select all simulations that are at most $3\sigma$ outliers from the mean of the posterior distribution. From the $\{\simp,\cdata\}$ pairs associated with these simulations, we then calculate the correlation matrix. Even this linear representation of our inference results is informative, but also rather complicated. While it is possible to understand many of the connections (e.g. the positive correlation between $\rho_{1\mathrm{kpc}}$ and $\text{std}_0(\tilde{w})$), it can be dubious to consider a single parameter or correlation in isolation, due to the strong degeneracies.

The top left of Figure~\ref{fig:corrs} shows correlations internally to $\simp$. The strongest correlation altogether is between the systemic transverse velocities $\tilde{u}$ and $\tilde{v}$, as discussed above. Moderately strong degeneracies also exist between the one of the two transverse velocities in relation to the matter density parameters and $\tilde{z}$. Correlations internally to $\cdata$ are shown in the bottom right, where the only strong degeneracies are between $\Xi$ and $s_\perp$, and between $\xi_\mathrm{turn}$ and $a_2$.

Perhaps the most interesting correlations, between $\simp$ and $\cdata$, are visible in the bottom left of Figure~\ref{fig:corrs}. These correlations indicate what the reduced data parameters are informative of; in other words, what $\simp$ parameters do the respective $\cdata$ parameters help constrain. We note that all correlations with $\tilde{w}$ are practically zero-valued; the prior over $\tilde{w}$ is already so narrow that other data cannot help constrain it further. The parameter $s_\perp$ has very weak correlations with most of $\simp$, except with $\tilde{v}$ and $\tilde{z}$, indicating that it is informative of the satellite's orbit. We see that $\xi_\mathrm{turn}$ is degenerate with the matter density parameters $\rho_{1\mathrm{kpc}}$ and $\gamma$, showing us that the precise shape of the velocity profile is informative of the satellite's self-gravity.

In the current version of this inference model, the only information we use about the sky-projected surface density profile is $\Xi$, the angle of semi-major axis. We do not include information about its ellipticity, meaning the relative size of semi-major and semi-minor spatial axes. We refrain from this in order to retain a more robust method, since this quantity is sensitive to deviations from the assumed initial conditions of isotropy and spherical symmetry. However, the correlation coefficient between ellipticity and $\tilde{z}$ has a value of 0.78, stronger than any $\tilde{z}$-correlation in Figure~\ref{fig:corrs}. This indicates that ellipticity, and potentially other features of the surface brightness profile, are highly informative, and that the inference model would benefit from using that information once it can faithfully reproduce NGC205's isophotal twist.

\section{Application to NGC205}
\label{sec:real_inference}

For the application to the real data of NGC205, we need to change our model of inference somewhat. In particular, the actual NGC205 velocity observations are distributed along a line, roughly following the semi-major axis of its twisting isophotes (see Figure~\ref{fig:real_NGC205}). In order to account for this difference, we modified our inference model in the following way. Since there is no information about the line-of-sight velocity field perpendicular to the semi-major axis, we remove the parameter $s_\perp$ from the reduced data vector $\cdata$, thus reducing its length from six to five. For the velocity data of the forward simulations, we use the same semi-major axis positions as of the real observations and match the data uncertainties. When compressing the real data, we ignore the twisting isophotes and take the velocity and position values, as presented in the bottom panel of Figure~\ref{fig:real_NGC205}, at face value. The semi-major axis angle ($\Xi$) is given by the angle of covariance of the surface brightness profile, according to the standard compressor method as described in Section~\ref{sec:compressor}.
Thus $\Xi$ falls in between the orientation angles of the innermost and outermost isophotes. For the compressed data vector target, the data presented in Section~\ref{sec:NGC205} gives the following values:
\begin{itemize}
    \item $\Xi = 1.994$~rad,
    \item $\xi_\mathrm{turn} = 3.05'$,
    \item $a_1 = 8.329~\kmsec$,
    \item $a_2 = -0.974~\kmsec$,
    \item $\text{std}_0(\tilde{w}) = 28.83~\kmsec$.
\end{itemize}

Our N-body simulations were not able to fully reproduce these data. Most importantly, we cannot find a simulation where the velocity profile is so strongly pronounced; in other words, we cannot produce a velocity S-shape with such a high amplitude (high $a_1$ and $a_2\simeq0~\kmsec$) and simultaneously such a short turn-around radius (small $\xi_\mathrm{turn}$). We can find simulations with the correct amplitude values, but only for $\xi_\mathrm{turn} \gtrsim 4'$.

We do not present any inferred posterior distribution in this section. Since we cannot fully reproduce the NGC205 observations quantitatively, the posterior would depend on the far tails of the Gaussians in the likelihood emulator. This is not what the inference framework is designed to do, making those results questionable.

Technical details aside, there is a significant systematic uncertainty associated with our modeling. We are able to reproduce the main features of NGC205 qualitatively, and come close in quantitative terms. In order to produce a good fit, we probably need to relax the simplifying assumptions that we make about the satellite's initial conditions, mainly those of perfect spherical symmetry and an isotropic velocity distribution. This is discussed further in Section~\ref{sec:discussion} below.

\section{Discussion}
\label{sec:discussion}

We qualitatively reproduce many of the features of NGC205, even starting from a spherically symmetric and isotropic state. If this initial state is at least close to correct and if the S-shaped velocity profile is predominantly produced by a tidal perturbation from M31, then that places strong constraints on NGC205's orbit. Firstly, the pericenter passage needs to have happened recently, in the past few 10~Myr. Secondly, the transverse velocity needs to be high, resulting in a strong but short-lived tidal impulse during the satellite's pericenter passage. Thirdly, in order to observe the S-shaped velocity profile, the line-of-sight vector needs to be close to parallel with the satellite's orbital plane. Fourthly, in order to get the correct sign on the S-like velocity profile, the satellite must have passed in front of M31. This places NGC205 on an orbit from the south-east towards the north-west (i.e. currently moving away from M31 on the sky).

This work's main precursor is \cite{Howley2008}, who performed an iterative search of NGC205's orbit using a genetic algorithm. They studied three different tracer population models (cold disk, warm disk, and spheroid), using fixed gravitational potentials for both host and satellite. We agree with their general conclusions: NGC205 likely has a large transverse velocity, is on a highly eccentric orbit, and thus probably in its first encounter with M31. In terms of our differences, they could not reproduce both an S-shaped velocity profile and the correct velocity dispersion with any single model, which we do. Furthermore, their best-fit orbits have a transverse velocity which is approaching M31, in the opposite direction from ours. Crucially, for the spatial depth relative to M31, \cite{Howley2008} searched the range of 2--76~kpc, thus excluding the orbit that we find to be a good fit.

There are two main shortcomings of our simulation models. Firstly, we can reproduce the ellipticity and angle of semi-major axis, but not NGC205's twisting isophotes. Secondly, the observed S-shaped velocity profile is somewhat more pronounced than what we can produce in our simulations, meaning a velocity S-shape with a short turn-around point ($\xi_\mathrm{turn} \simeq 3'$) and high amplitude ($a_1\simeq8~\kmsec$). It seems possible to reconcile these shortcomings by relaxing the simplifying assumptions that we make for the satellite's initial conditions. In Section~\ref{sec:internal_rotation}, we show with a simple example that adding a small or moderate amount of internal rotation can in fact give rise to twisting isophotes, and also make the velocity S-shape more pronounced.

In order to improve our modeling and fully exploit the data, we plan to include internal rotation in future version of our inference model. Doing so increases the dimensionality of our problem, both in terms of the simulation input parameters ($\simp$), as well as the number of informative reduced data parameters ($\cdata$), and will require running a larger number of simulations. This inference problem is already computationally expensive, and N-body simulations are the dominant cost.
\update{In order to ameliorate this issue, we plan to test using N-body integration with basis function expansions \citep{Petersen2022}. In this N-body simulation framework, the phase-space density of particles and the corresponding gravitational potential are described by a basis function expansion.}
Calculating the potential in this manner can speed up simulations, where the computational time scales like $\mathcal{O}(N)$, rather than $\mathcal{O}(N^2)$ or $\mathcal{O}(N\log N)$ for direct and tree-based algorithms. \update{Before applying this framework, we need to test it thoroughly by comparing with standard N-body simulations, in order to ensure accurate results.} Another improvement is to demote $\tilde{w}$ to a nuisance parameter or even a fixed value, since our modeling does not provide any additional constraining power to this parameter.

We saw in our tests on mock data that the mean velocity field along the semi-minor axis is informative, in particular of the satellite's orbit. In addition, internal rotation can affect the observed velocity field in non-trivial ways, for example by breaking symmetry across the semi-major axis. For these reasons, it would be impactful to revisit NGC205 to take more velocity data, in particular to make observations that are more spatially uniform, rather than just distributed along the semi-major axis. A systemic proper motion measurement would also help test our suggested orbit and mechanism to produce the S-shaped velocity profile.

Our method uses the example of NGC205 as a testbed, but it also has broader implications. There are other dwarf galaxy satellites that have been significantly perturbed by a close passage with their host galaxy: for example, NGC770 (a satellite of NGC772) exhibits an S-shaped velocity profile similar to NGC205 \citep{Geha2005}, and our own Milky Way is host to perturbed systems such as Crater II \citep{Fu2019,Borukhovetskaya2022} and the Sagittarius dwarf galaxy \citep{Ibata1994}. We have provided a possible framework for performing precision inference on such heavily perturbed systems, and have demonstrated that their time-varying dynamics can be highly informative, for example of their orbit and total matter density profile. We can do so even in the lack of proper motion information, which makes this framework well suited for extra-galactic systems. 

\section{Conclusion}
\label{sec:conclusion}

We have developed a novel approach to precisely infer the orbit and mass density profile of a tidally perturbed dwarf galaxy, in a Bayesian framework with forward simulation and likelihood emulators. As a testbed for this method, we consider the case of NGC205, a satellite of M31 with S-like shapes in its surface brightness profile, and in its line-of-sight velocity as a function of semi-major axis. We show that NGC205's observed line-of-sight velocity field can be qualitatively reproduced even if NGC205 was in an isotropic and spherically symmetric state before its most recent pericenter passage. We simultaneously reproduce the broad features of its surface brightness profile, in terms of its overall ellipticity and angle of semi-major axis. We test our method on mock data and demonstrate that we can precisely retrieve its orbit, as well as its initial mass density profile, even with decent constraining power on the mass density's inner slope.

For inference on the actual NGC205, our method is hampered by issues with data and modeling. The currently available velocity data only covers a narrow line along its semi-major axis. Our mock data results shows that having more uniformly sampled velocity measurements is informative especially of the satellite's orbit. Our current model does not replicate the twisting surface brightness profile of NGC205. However, we show with simulations that this feature can be reproduced by introducing rotation to the satellite's initial conditions. We plan to extend our model to include this effect in future work.

The inference method we have developed demonstrates that the internal time-varying dynamics of tidally perturbed dwarf galaxies can be highly informative, and even allows for precise dynamical mass measurements. In the context of current and near future observations, this inference framework can be used to understand and analyze individual systems, potentially as a probe of the dynamical properties of dark matter.


\section*{Acknowledgments}

We wish to express our gratitude towards Matthew Ho, Marla Geha, and Michael S. Petersen for insightful and productive discussions.
This research utilized the Sunrise HPC facility supported by the Technical Division at the Department of Physics, Stockholm University.
AW is supported by the European Union's Horizon 2020 research and innovation program, under the Marie Skłodowska-Curie grant agreement number 101106028. KVJ is supported by Simons Foundation grant 1018465.
This work made use of {\tt Numpy}~\citep{Harris_2020}, {\tt SciPy}~\citep{Virtanen:2019joe}, {\tt matplotlib}~\citep{HunterMatplotlib}.

\bibliography{refs}{}
\bibliographystyle{aasjournal}



\appendix

\section{M31 gravitational potential}
\label{app:M31_pot}

For the gravitational potential of M31, we use results from \cite{Zhang2024}. They use observation of stars, emission-line objects, and clusters located in M31's bulge, disk, and halo, using data from LAMOST DR9 \citep{Luo2015}, DESI \citep{Dey2022}, and various literature sources.

They fit a potential model consisting of a Hernquist bulge, a Miyamoto-Nagai disk, and an NFW halo. Their respective gravitational potentials, with subscripts $\{b,d,h\}$, take the form
\begin{equation}
\begin{split}
    \Phi_b(r) & = -\frac{GM_b}{r+q_b}, \\
    \Phi_d(R,z) & = -\frac{GM_d}{\sqrt{R^2+\Big(a_d^2+\sqrt{z^2+b_d^2}\Big)^2}} \\
    \Phi_h(r) & = -\frac{GM_h \ln (1+rc_h/r_{v,h})}{r\big[\ln(1+c_h)-c_h/(1+c_h)\big]},
\end{split},
\end{equation}
where $G$ is the gravitational constant, $r$ is radius, and $R$ and $z$ are radius and height in cylindrical coordinates.

The gravitational potential model we use in this work is given by the median values of the posterior density distribution inferred by \cite{Zhang2024}. They are
\begin{equation}
    \begin{split}
        \text{log}_{10}(M_b/\Msun) & = 10.57_{-0.28}^{+0.17}, \\
        q_b & = 840_{-270}^{+460}~\pc, \\
        \text{log}_{10}(M_d/\Msun) & = 10.86_{-0.36}^{0.20}, \\
        a_d & = 8770_{-3790}^{+2140}~\pc, \\
        b_d & = 550_{-350}^{+1830}~\pc, \\
        M_h & = 1.14_{-0.35}^{+0.51} \times 10^{12}~\Msun, \\
        r_{v,h} & = 220 \pm 25~\kpc, \\
        \text{log}_{10}(c_h/\Msun) & = 0.94_{-0.35}^{+0.25}, \\
    \end{split}
\end{equation}
where the plus and minus values denoting the 16th and 84th percentiles. The gravitational potential of M31 is uncertain by roughly 20~\%.

\section{Test of steady state initialization}
\label{app:eq_test}

In order to test that our simulations are indeed initialized in a steady state, we perform a test where a dwarf galaxy is evolved in isolation (i.e. no external tidal field). We use the same initial conditions as for our fiducial simulation, although we only include massive particles.

\begin{figure}
    \centering
    \includegraphics[width=.96\columnwidth]{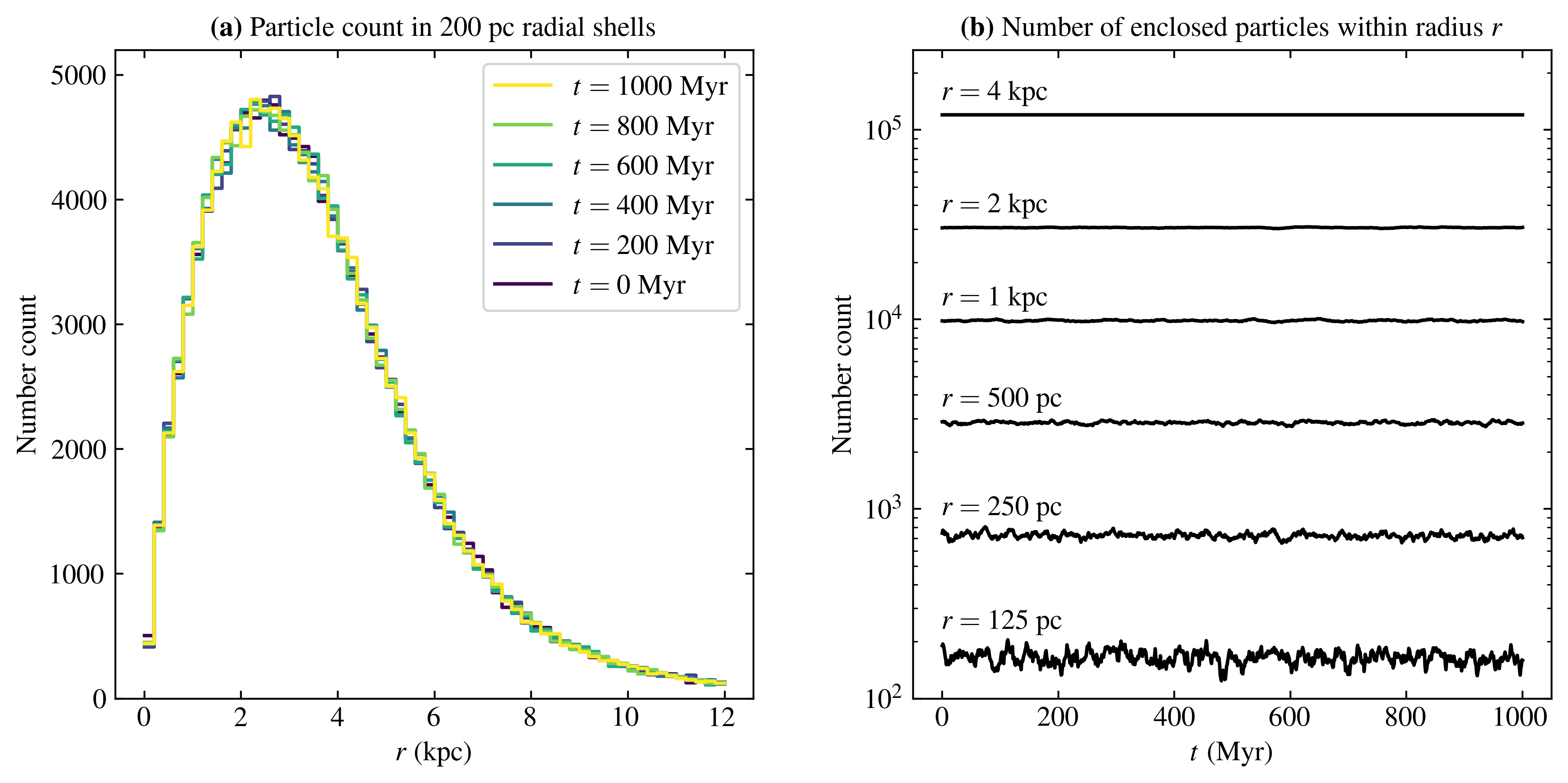}
    \caption{
    Evolution of a simulation without any external tidal field, as a test of our steady state initialization. Here we have used the same initial conditions as for our fiducial simulation, although only including the massive particles. The time $t=0~\Myr$ correspond to the unevolved, initial state of the simulation. The left panel shows the number of particles within radial shells with a width of 200~pc. In the right panel, we show the number of particles enclosed within six different radii, logarithmically spaced between 125~pc and 4~kpc, as a function of time. 
    }
    \label{fig:eq_test}
\end{figure}

In Figure~\ref{fig:eq_test}, we show how the number count radial profile evolves over time. The system is evolved for 1000~Myr, where the time $t=0~\Myr$ correspond to its initial state. For the innermost radii, the number of enclosed particles are somewhat noisy, but this is consistent with statistical Poisson noise. 
We conclude that the steady state initialization of the N-body system works well.

\section{Fiducial simulation with internal rotation}
\label{app:internal_rotation}

\begin{figure}
    \centering
    \includegraphics[width=1\textwidth]{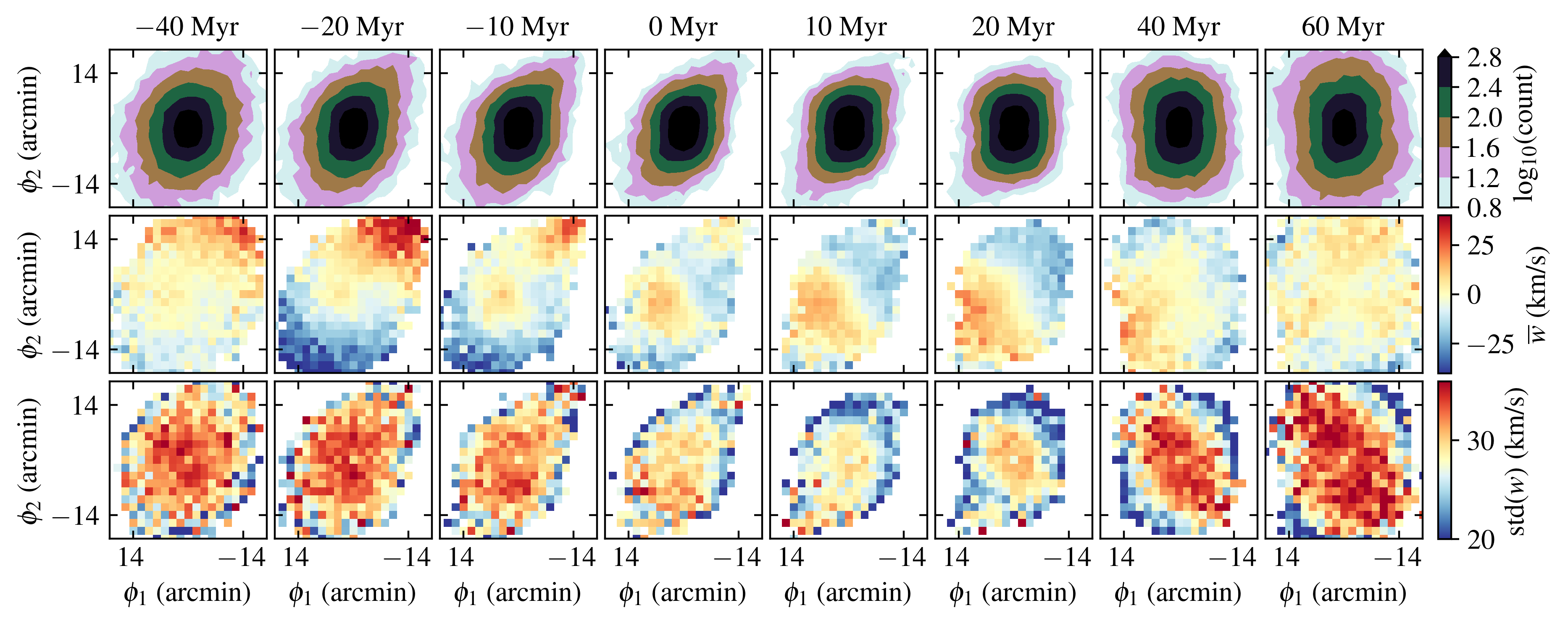}
    \caption{Same as Figure~\ref{fig:sim_obs_proj}, showing the surface density and velocity fields of the fiducial simulation, although here modified such that the tracer population is initialized with a moderate amount of internal rotation.}
    \label{fig:sim_obs_proj_with_rot}
\end{figure}

In Figure~\ref{fig:sim_obs_proj_with_rot}, we show the observable fields of our fiducial simulation, analogous to Figure~\ref{fig:sim_obs_proj}, but where the tracer population is initialized with a moderate amount of internal rotation. We used a rotation axis which is parallel to the observer's line-of-sight. See Section~\ref{sec:internal_rotation} for further details.

\end{document}